\documentclass[preprint,showkeys,showpacs]{revtex4-1}
\usepackage{amssymb}
\usepackage{amsmath}
\usepackage{graphicx}
\renewcommand{\figurename}{Fig.}

\begin{document} 
\title{Diffuse relaxation approximation in a heated Fermi system}
\author{S.V. Lukyanov}
\affiliation{Institute for Nuclear Research, 03680 Kyiv, Ukraine}

\begin{abstract}
An expression for the two-particle relaxation time of collective excitations
on a distorted Fermi surface in the diffusion approach to kinetic theory is obtained.
The general case of momentum-dependent diffusion and drift coefficients is considered.
The temperature dependence of the obtained expression is established. 
\end{abstract}


\maketitle

\section{Introduction}

It is convenient to study relaxation processes in a multiparticle Fermi system using quantum kinetic theory 
\cite{LiPi.bp2.1980,BePe.b.1991}. In the kinetic approximation, the system is described by the Landau-Vlasov 
equation for the Wigner distribution function in the phase space of coordinates and momenta. 
The advantage of this approach is the ability to simply describe the average values of such quantities as 
nucleon density, flux density, pressure, etc. for the case of a quantum Fermi system. However, difficulties 
of a different nature arise here due to the presence of the nine-dimensional collision integral 
in the right-hand side of the kinetic equation \cite{Be.ZP.1978,KoLuPlSh.PRC.1998}. 

To solve the problem with the collision integral, various methods are used to simplify it 
\cite{KoLuPlSh.PRC.1998,KoSh.PR.2004}. In particular, one of such methods is the diffuse approximation 
\cite{LiPi.bp2.1980}. Within the framework of this approximation, Wolschin's paper \cite{Wo.PRL.1982} considered 
a schematic model describing the equilibrium state in the Fermi system with finite dimensions. 
The master equation for single-particle filling states has been transformed into a nonlinear partial 
differential equation using the Pauli principle. Within the framework of this model, the author obtained 
an analytical solution in the simplified case of constant transport coefficients, which were introduced as moments 
of scattering probability. However, the numerical values of the diffusion and drift coefficients were used 
without any justification.

In the works \cite{KoLu.UPJ.2014,KoLu.IJMP.2015} it was demonstrated that the kinetic Landau-Vlasov equation 
with the collision integral can be reduced to the form of the diffusion equation in momentum space. 
Explicit expressions are also obtained for the diffusion and drift coefficients
for the Fermi system in which relaxation processes occur on a deformed Fermi surface.
It turned out that for the correct calculation of the kinetic coefficients it is not enough 
to simply assume the isotropic character of the scattering probability in the collision integral. 
To describe the interparticle interaction, one should assume forward scattering of particles. 
Therefore, a Gaussian-type potential with a limited radius of action was chosen 
and the dependences of these coefficients on momenta and temperature were calculated. 
In the approximation of the constant diffusion and drift coefficients, an expression is obtained 
for the relaxation time of collective multipole excitations. 

In this work, we calculated the two-particle relaxation time in the general case 
of the momentum-dependent transport diffusion and drift coefficients.
The dependence of the obtained expression for the renormalized relaxation time 
on temperature is investigated.

\section{Diffusion approximation for the kinetic equation}

Consider the kinetic equation with the collision integral 
\begin{equation}
\frac{\partial f(\mathbf{r},\mathbf{p},t)}{\partial t}+\hat{L}f(\mathbf{r},\mathbf{p},t) 
= \mathrm{St}\{f\},
\label{kineq}
\end{equation}
were $f(\mathbf{r},\mathbf{p},t)$ is the Wigner distribution function in the phase space, 
$\mathrm{St}\{f\}$ is the collision integral, and the operator $\hat{L}$ is given by the expression
\begin{equation}
\hat{L}=\frac{1}{m}\mathbf{p}\cdot\mathbf{\nabla}_\mathbf{r}
-\left(\mathbf{\nabla}_\mathbf{r}U\right)\cdot\mathbf{\nabla}_\mathbf{p}.
\label{operator_l}
\end{equation}
In the general case, the one-particle potential $U$ includes the self-consistent and external fields.
We choose the collision integral in the form \cite{LiPi.bp2.1980,KoLu.IJMP.2015}
\begin{equation}
\mathrm{St}\{f\}=\int\frac{gd\mathbf{p}_{3}}{(2\pi\hbar)^{3}}\ 
\left[ W_{3\rightarrow 1}(\mathbf{p}_{1},\mathbf{p}_{3})\tilde{f}(\mathbf{p}_{1})
f(\mathbf{p}_{3})
-W_{1\rightarrow 3}(\mathbf{p}_{1},\mathbf{p}_{3})f(\mathbf{p}_{1})
\tilde{f}(\mathbf{p}_{3})\right],  
\label{st1}
\end{equation}
where $f(\mathbf{r},\mathbf{p}_j,t)\equiv f(\mathbf{p}_j)\equiv f_j$ and $f_1\equiv f$,
$g=4$ is the spin-isospin degeneracy factor, $\tilde{f}(\mathbf{p}_j)=1-f(\mathbf{p}_j)$. 
The gain and loss terms $W_{3\leftrightarrows 1}(\mathbf{p}_{1},\mathbf{p}_{3})$ in Eq. 
(\ref{st1}) have the form \cite{KoLu.IJMP.2015}
\begin{eqnarray}
W_{3\rightarrow 1}(\mathbf{p}_{1},\mathbf{p}_{3}) &\equiv &W(\mathbf{p}_{1},\mathbf{p}_{3})
=\int \frac{gd\mathbf{p}_{2}d\mathbf{p}_{4}}{(2\pi \hbar )^{3}}\ 
\mathcal{W}(\{\mathbf{p}_{j}\})\tilde{f}(\mathbf{p}_{2})f(\mathbf{p}_{4})
\delta (\Delta \varepsilon )\delta (\Delta \mathbf{p}),  
\label{gains}
\\
W_{1\rightarrow 3}(\mathbf{p}_{1},\mathbf{p}_{3}) &\equiv &\widetilde{W}
(\mathbf{p}_{1},\mathbf{p}_{3})
=\int \frac{gd\mathbf{p}_{2}d\mathbf{p}_{4}}
{(2\pi \hbar )^{3}}\ \mathcal{W}(\{\mathbf{p}_{j}\})f(\mathbf{p}_{2})
\tilde{f}(\mathbf{p}_{4})\delta (\Delta \varepsilon )\delta (\Delta \mathbf{p}), 
\label{losses}
\end{eqnarray}
where $\mathcal{W}(\{\mathbf{p}_{j}\})$ is the probability of two-particle collisions,
$\Delta \mathbf{p}=\mathbf{p}_{1}+\mathbf{p}_{2}-\mathbf{p}_{3}-\mathbf{p}_{4}$,
$\Delta \varepsilon =\varepsilon _{1}+\varepsilon _{2}-\varepsilon _{3}-\varepsilon_{4}$. 
Here $\varepsilon_j=p_j^2/2m+U(r_j)$ is one-particle energy.
The quantity $W_{i\rightarrow j}(\mathbf{p}_{1},\mathbf{p}_{3})$ is the probability scattering 
with the transition of a particle from the state $\mathbf{p}_i$ to the state $\mathbf{p}_j$ 
surrounded by particles of the medium.

It should be noted here that the spin-averaged probability of two-particle collisions $\mathcal{W}(\{\mathbf{p}_{j}\})$
in Eqs (\ref{gains}), (\ref{losses}) can be expressed in terms of the cross section of scattering 
in nuclear matter $d\sigma /d\Omega$ as follows
\begin{equation}
\mathcal{W}(\{\mathbf{p}_{j}\})=\frac{2(2\pi \hbar )^{3}}{m^{2}}
\frac{d\sigma }{d\Omega }(\{\mathbf{p}_{j}\}).  
\label{wdsdo}
\end{equation}
In the case of elastic collisions the scattering cross section $d\sigma /d\Omega$ depends only 
on the square of the modulus of the momentum transferred $\mathbf{s}$.
To ensure the smallness of the transferred momentum $\mathbf{s}$ we will use the expression 
of the Gaussian type as the differential cross section $d\sigma /d\Omega$ \cite{Dav.b.65}
\begin{equation}
\frac{d\sigma (\{\mathbf{p}_{j}\})}{d\Omega }
=\frac{\pi m^{2}r_{0}^{6}v_{0}^{2}}{2\hbar^{4}}
\exp \left(-4\mathbf{s}^{2}r_{0}^{2}/\hbar^{2}\right),  
\label{pot}
\end{equation}
where $r_{0}$ and $v_{0}$ are free parameters. Thus, the main contribution to the scattering amplitude 
will come from transitions corresponding to a small transferred momentum: $|\mathbf{p}_{1}-\mathbf{p}_{3}|\ \ll p_{F}$, 
where $p_{F}$ is the Fermi momentum. Introducing variables
$$
\mathbf{s}=\mathbf{p}_{3}-\mathbf{p}_{1}\quad \text{ta}\quad \mathbf{P}
=\frac{1}{2}(\mathbf{p}_{1}+\mathbf{p}_{3})=\mathbf{p}_{1}+\frac{\mathbf{s}}{2},
$$
we apply the appropriate expansions in small momentum transfer $\mathbf{s}$:
\begin{equation}
f(\mathbf{p}_{3})=f(\mathbf{p}_{1}+\mathbf{s})\approx f(\mathbf{p}_{1})
+s_{\nu }\nabla _{p_{1},\nu }\ f(\mathbf{p_{1}})+\frac{1}{2}\ s_{\nu}s_{\mu}
\nabla _{p_{1},\nu }\nabla _{p_{1},\mu }\ f(\mathbf{p_{1}}),
\label{f3}
\end{equation}
\begin{eqnarray}
W(\mathbf{p}_{1},\mathbf{p}_{3}) &=&W(\mathbf{P},\mathbf{s})  \nonumber \\
&\approx &W(\mathbf{p}_{1},\mathbf{s})+\frac{1}{2}\ s_{\nu }\nabla
_{p_{1},\nu }W(\mathbf{p}_{1},\mathbf{s})+\frac{1}{8}\ s_{\nu }s_{\mu
}\nabla _{p_{1},\nu }\nabla _{p_{1},\mu }W(\mathbf{p}_{1},\mathbf{s}),
\label{w13}
\end{eqnarray}
and
\begin{eqnarray}
\widetilde{W}(\mathbf{p}_{1},\mathbf{p}_{3}) &=&\widetilde{W}(\mathbf{P},
\mathbf{s})  \nonumber \\
&\approx &\widetilde{W}(\mathbf{p}_{1},\mathbf{s})+\frac{1}{2}\ s_{\nu
}\nabla _{p_{1},\nu }\widetilde{W}(\mathbf{p}_{1},\mathbf{s})+\frac{1}{8}\
s_{\nu }s_{\mu }\nabla _{p_{1},\nu }\nabla _{p_{1},\mu }
\widetilde{W}(\mathbf{p}_{1},\mathbf{s}).  
\label{w31}
\end{eqnarray}

After the corresponding transformations described in the paper \cite{KoLu.IJMP.2015}, 
we obtain the collision integral in the diffusion approximation
\begin{equation}
\mathrm{St}\{f\} = 
- \nabla_{p,\nu} \left[ K_p(\mathbf{p}) f(\mathbf{p})\tilde{f}(\mathbf{p}) \frac{p_\nu}{m} 
+ f(\mathbf{p})^2 \nabla_{p,\nu} D_p(\mathbf{p})  \right] 
+ \nabla_p^2\left[f(\mathbf{p}) D_p(\mathbf{p})\right],
\label{stf}
\end{equation}
where $D_{p}(\mathbf{p})$ and $K_{p}(\mathbf{p})$ are the diffusion and drift coefficients 
in momentum space, respectively.
Both kinetic coefficients $D_{p}(\mathbf{p})$ and $K_{p}(\mathbf{p})$ satisfy the following relations
\begin{equation}
D_{p}(\mathbf{p})=\frac{1}{6}\int \frac{gd\mathbf{s}}{(2\pi \hbar )^{3}}\ 
s^{2}\ W(\mathbf{p},\mathbf{s})  
\label{dpdef1}
\end{equation}
and
\begin{equation}
K_{p}(\mathbf{p})\nabla _{p,\nu }\varepsilon _{p}=\nabla _{p,\nu }D_{p}(\mathbf{p})-
\int \frac{gd\mathbf{s}}{(2\pi \hbar )^{3}}\ s_\nu\ W(\mathbf{p},\mathbf{s}).  
\label{kpdef1}
\end{equation}

To analyze the dependence of the collision integral on the multipolarity of the Fermi surface deformation 
we will consider a small deviation $\delta f$ of the distribution function from its equilibrium value $f_{eq}$, 
i.e.
\begin{equation}
\delta f(\mathbf{p},t)=f(\mathbf{p},t)-f_{eq}(\mathbf{p}).
\label{df}
\end{equation}
Then, for the linearized distribution function (\ref{df}), the collision integral (\ref{stf}) 
is also linearized up to quadratic terms that are insignificant in smallness
\begin{equation}
\mathrm{St}\{f\}\simeq \mathrm{St}\{f_{eq}\} + \mathrm{St}\{\delta f\}.
\label{stflin}
\end{equation}

For the equilibrium distribution function, the collision integral is identically equal to zero, 
therefore, substituting Eq. (\ref{df}) into Eq. (\ref{stf}) we have
\begin{eqnarray}
f_{eq}(\mathbf{p})(1-f_{eq}(\mathbf{p})) \nabla_{p,\nu}\displaystyle \left[ K_p(\mathbf{p}) 
\frac{p_\nu}{m} \right] + K_p(\mathbf{p}) \frac{p_\nu}{m} (1-2f_{eq}(\mathbf{p})) 
\nabla_{p,\nu} f_{eq}(\mathbf{p}) \nonumber \\ 
- D_p(\mathbf{p}) \nabla^2_{p} f_{eq}(\mathbf{p}) - 2(1-f_{eq}(\mathbf{p})) 
\nabla_{p,\nu}f_{eq}(\mathbf{p}) \cdot \nabla_{p,\nu} D_p(\mathbf{p}) \nonumber \\
- f_{eq}(\mathbf{p}) (1-f_{eq}(\mathbf{p})) \nabla^2_{p} D_p(\mathbf{p}) =0.
\label{stfeq1}
\end{eqnarray}

We will assume that the equilibrium distribution function has spherical symmetry in momentum space
$f_{eq}(\mathbf{p})=f_{eq}(p)$. Then Eq. (\ref{stfeq1}) takes the form
\begin{eqnarray}
f_{eq}(p)(1-f_{eq}(p)) \frac{\partial}{\partial p_\nu} \left[ K_p(p) \frac{p_\nu}{m} \right] 
+ K_p(p) \frac{p_\nu}{m} (1-2f_{eq}(p)) \frac{\partial f_{eq}(p)}{\partial p_\nu} \nonumber \\
- D_p(p) \frac{\partial^2 f_{eq}(p)}{\partial p^2} 
-\frac{2 D_p(p)}{p} \frac{\partial f_{eq}(p)}{\partial p} 
- 2 (1-f_{eq}(p)) \frac{\partial f_{eq}(p)}{\partial p_\nu} \frac{\partial D_p(p)}{\partial p_\nu}  
\nonumber \\
- f_{eq}(p) (1-f_{eq}(p)) \left( \frac{\partial^2 D_p(p)} {\partial p^2} 
+ \frac{2}{p} \frac{\partial D_p(p)}{\partial p} \right)=0.
\label{stfeq2}
\end{eqnarray}

For a spherically symmetric equilibrium Fermi distribution function
\begin{equation}
f_{eq}(p)=\left[1+\exp\left(\frac{p^2/2m-\varepsilon_F}{T}\right)\right]^{-1},
\label{feqfermi}
\end{equation}
where $\varepsilon_F$ is the Fermi energy and $T$ is the temperature parameter, 
the following relations will be valid
\begin{equation}
\frac{\partial f_{eq}(p)}{\partial p} = - \frac{p}{mT} f_{eq}(p)(1-f_{eq}(p)),
\label{d1fermi}
\end{equation}
\begin{equation}
\frac{\partial^2 f_{eq}(p)}{\partial p^2} = - \frac{1}{mT} \left[1-\frac{p^2}{mT}(1-2f_{eq}(p))\right]
f_{eq}(p)(1-f_{eq}(p)).
\label{d2fermi}
\end{equation}
Taking into account the relations (\ref{d1fermi}) and (\ref{d2fermi}) we rewrite Eq. (\ref{stfeq2}) as follows
\begin{eqnarray}
f_{eq}(p)(1-f_{eq}(p)) \left\{ \frac{p}{m} \frac{\partial K_p(p)}{\partial p} + 3\frac{K_p(p)}{m} 
- \frac{K_p(p)}{m} \frac{p^2}{mT} (1-2f_{eq}(p)) \right. 
\nonumber \\ 
+ \left.\frac{D_p(p)}{mT} \left[1-\frac{p^2}{mT}(1-2f_{eq}(p))\right]\right.
+ \frac{2 D_p(p)}{mT} + 2 (1-f_{eq}(p)) \frac{p}{mT}\frac{\partial D_p(p)}{\partial p}  
\nonumber \\
- \left. \left(\frac{\partial^2 D_p(p)} {\partial p^2} + \frac{2}{p} \frac{\partial D_p(p)}
{\partial p} \right)\right\}=0.
\label{stfeq3}
\end{eqnarray}
The expression in curly braces is zero. After the reductions we get
\begin{eqnarray}
\frac{1}{m}\left[ 3 - \frac{p^2}{mT} (1-2f_{eq}(p)) \right] \left(K_p(p) + \frac{D_p(p)}{T}\right)
\nonumber \\
+ \frac{p}{m} \frac{\partial K_p(p)}{\partial p} 
- \frac{2}{p} \left[1 - \frac{p^2}{mT}(1-f_{eq}(p))\right] \frac{\partial D_p(p)}{\partial p}  
- \frac{\partial^2 D_p(p)} {\partial p^2} =0.
\label{stfeq4}
\end{eqnarray}
This equation determines the temperature parameter $T$ for arbitrary values of the kinetic diffusion 
and drift coefficients.

In the case of constant kinetic coefficients $D_p=\mathrm{const}$ and $K_p=\mathrm{const}$ 
from Eq. (\ref{stfeq4}) we get
\begin{equation}
\left[ 3 - \frac{p^2}{mT} (1-2f_{eq}(p)) \right] \left(K_p + \frac{D_p}{T}\right)=0.
\label{stfeq5}
\end{equation}
As you can see from Eq. (\ref{stfeq5}) the relationship between the temperature parameter and constant 
values of the kinetic coefficients is fulfilled in almost the entire momentum space
\begin{equation}
T=-\frac{D_p}{K_p}.
\label{temp}
\end{equation}

Let us expand a small deviation $\delta f$ of the distribution function from the equilibrium
in a series in spherical functions
\begin{equation}
\delta f=-\frac{\partial f_{eq}}{\partial \varepsilon} \sum_{lm}\nu_{lm}(\mathrm{r},t)Y_{lm}(\Omega_p).
\label{dfsumynm}
\end{equation}

We define the relaxation time $\tau_l$ for the distorted Fermi surface with multipolarity $l$ as follows
\begin{equation}
\tau_l^{-1} =-\frac{\displaystyle\int d\mathrm{p}\ \mathrm{St}\{\delta f\}Y_{lm}(\Omega_p)}
{\displaystyle\int d\mathrm{p}\ \delta f\ Y_{lm}(\Omega_p)}.
\label{taul}
\end{equation}

We represent the linearized part of the collision integral $\mathrm{St}\{\delta f\}\equiv \delta\mathrm{St}$ 
(\ref{stflin}) in the form
\begin{eqnarray}
\delta\mathrm{St}=
&-&\nabla_{p,\nu}\left[K_p(\mathbf{p}) (1-2f_{eq}(\mathbf{p}))\delta f(\mathbf{p}) \frac{p_\nu}{m}
+ 2 f_{eq}(\mathbf{p}) \delta f(\mathbf{p}) \nabla_{p,\nu} D_p(\mathbf{p})\right] 
\nonumber \\
&+& \nabla_p^2\left[\delta f(\mathbf{p}) D_p(\mathbf{p})\right].
\label{dstf}
\end{eqnarray}
For the spherically symmetric distribution (\ref{feqfermi}), we obtain
\begin{eqnarray}
\delta\mathrm{St}=
-K_p(p)\nabla_{p,\nu}\left[(1-2f_{eq}(p))\delta f(p) \frac{p_\nu}{m}\right]
-(1-2f_{eq}(p))\delta f(p) \frac{p}{m}\nabla_{p}K_{p}(p) 
\nonumber \\
-2\delta f(p)\ \nabla_{p}f_{eq}(p)\cdot\nabla_{p} D_p(p)
+2(1-f_{eq}(p)) \nabla_{p}\delta f(p) \cdot \nabla_{p} D_p(p)
\nonumber \\
+(1-2 f_{eq}(p)) \delta f(p) \nabla_{p}^2 D_p(p)
+ D_p(p) \nabla_p^2 \delta f(p).
\label{dstf1}
\end{eqnarray}

We calculate the $l$-th moment of the obtained linearized collision integral
\begin{eqnarray}
\int d\vec{p}\ Y_{lm}(\Omega_p) \delta\mathrm{St}
=&-& \int d\vec{p}\ Y_{lm}(\Omega_p) K_p(p) \nabla_{p,\nu} \left[(1-2f_{eq}(p))\delta f(p) 
\frac{p_\nu}{m}\right]
\nonumber \\
&-& \int d\vec{p}\ Y_{lm}(\Omega_p) (1-2f_{eq}(p))\delta f(p) \frac{p}{m} \nabla_{p} K_p(p)
\nonumber \\
&-& 2 \int d\vec{p}\ Y_{lm}(\Omega_p)\delta f(p)\ \nabla_{p}f_{eq}\cdot\nabla_{p} D_p(p)
\nonumber \\
&+& 2 \int d\vec{p}\ Y_{lm}(\Omega_p) (1-f_{eq}(p)) \nabla_{p}\delta f(p) \cdot \nabla_{p} D_p(p)
\nonumber \\
&+& \int d\vec{p}\ Y_{lm}(\Omega_p) (1-2 f_{eq}(p)) \delta f(p) \nabla_{p}^2 D_p(p)
\nonumber \\
&+& \int d\vec{p}\ Y_{lm}(\Omega_p) D_p(p) \nabla_p^2 \delta f(p)
\nonumber \\
&=&-J_1-J_2-2J_3+2J_4+J_5+J_6,
\label{intdstfl}
\end{eqnarray}
where the corresponding integrals are denoted by $J_n$, $n=1, 2, ..., 6$. 
Their calculation is given in the Appendix. 
As we can see, the integrals $J_1$ and $J_2$ cancel each other, and the integral $J_3$ cancels out with 
the second term in the expression for the integral $J_4$.
So, after substitution and final abbreviations we have
\begin{equation}
\int d\vec{p}\ Y_{lm}(\Omega_p) \delta\mathrm{St}
=l(l+1)m\nu_{lm}\int_0^\infty dp\ \frac{\partial f_{eq}(p)}{\partial p} \frac{D_p(p)}{p}.
\label{intdstfl1}
\end{equation}

Substituting Eq. (\ref{intdstfl1}) into Eq. (\ref{taul}) and noticing that
\begin{equation}
\int d\vec{p}\ Y_{lm}(\Omega_p) \delta f(p) = 
- m\nu_{lm} \int_0^\infty dp\ p\frac{\partial f_{eq}(p)}{\partial p},
\end{equation}
we end up with
\begin{equation}
\tau_l^{-1}=l(l+1)\ 
\frac{\displaystyle\int_0^\infty dp\ \frac{\partial f_{eq}(p)}{\partial p}\frac{D_p(p)}{p}}
{\displaystyle\int_0^\infty dp\ p\ \frac{\partial f_{eq}(p)}{\partial p}}.
\label{taul1}
\end{equation}
It should be noted here that the expression (\ref{taul1}) is accurate and obtained without any simplifications.

In the case of low temperatures $T\ll E_F$, the equilibrium distribution function is close to 
the stepwise distribution in momentum space; therefore, its derivative has a delta-like form
$\frac{\displaystyle\partial f_{eq}}{\displaystyle\partial p} \sim -\delta(p-p_F)$. 
Taking this fact into account, the expression (\ref{taul1}) in the low-temperature approximation 
takes the form \cite{KoLu.UPJ.2014}
\begin{equation}
\tau_l^{-1} = l(l+1)\ \frac{D_p(p_F)}{p^2_F}.
\label{taulf}
\end{equation}
It should be noted here that the expression (\ref{taulf}) coincides with the expression for 
the two-particle relaxation time, which was obtained by us in \cite{KoLu.UPJ.2014}
in the approximation of constant diffusion and drift coefficients.
Consequently, the approximation of constant diffusion and drift coefficients does not exclude
their dependence on temperature.

Looking at Eqs. (\ref{taul1}) and (\ref{taulf}), it should be noted that the two-particle relaxation 
time is determined only through the diffusion coefficient $D_p(p)$.

\section{Numerical calculations}

For clarity of the results obtained, we will numerically calculate 
the dependences of the obtained expressions on the temperature parameter $T$.
When calculating the two-particle relaxation time, we will use the expression obtained in our work 
\cite{KoLu.IJMP.2015} as the diffusion coefficient in momentum space 
\begin{eqnarray}
D_{p}(p) &\approx & \frac{g^{2}mr_{0}^{6}v_{0}^{2}}{6 \hbar^{7}}
\int_{0}^{\infty }dk\ k^{5}\int_{-1}^{1}dx\int_{-1}^{1}dy\ \exp \left[
-8k^{2}r_{0}^{2}(1-xy)/\hbar ^{2}\right]  \nonumber \\
&\times &\left\{ (1-xy)\ I_0\left( 8k^{2}r_{0}^{2}\sqrt{1-x^{2}}\sqrt{1-y^{2}}
/\hbar ^{2}\right) -\sqrt{1-x^{2}}\sqrt{1-y^{2}}\right.  \nonumber \\
&\times &\left. I_1\left( 8k^{2}r_{0}^{2}\sqrt{1-x^{2}}
\sqrt{1-y^{2}}/\hbar ^{2}\right) \right\} \tilde{f}\left( \sqrt{k^{2}+p^{2}+2kpx}
\right)  \nonumber \\
&\times &f\left( \sqrt{k^{2}+p^{2}+2kpy}\right),  
\label{dp8x}
\end{eqnarray}
where $I_0(z)$ and $I_1(z)$ are modified Bessel functions of the first kind. For free parameters 
of internucleon interaction, we choose the following values \cite{KoLu.UPJ.2014,KoLu.IJMP.2015}: 
$r_{0}=0.8$ \textrm{fm} and $v_{0}=-33$ \textrm{MeV}. 
It is these values of the parameters that provide a reasonable value of the in-medium nucleon-nucleon
cross-section $\sigma _{\mathrm{tot}}\simeq 20$ \textrm{mb}. 
For the Fermi energy we choose the typical value $\varepsilon_{F}=37$ \textrm{MeV}.

As can be seen from Eqs. (\ref{taul1}) and (\ref{taulf}), the calculation results, up to the coefficient 
$l(l+1)$, will be identical for all multipolarities of the Fermi surface deformation.
Therefore, below we will consider the renormalized relaxation time $l(l+1)\tau_l$.

\begin{figure}[tbp]
\begin{center}
\includegraphics[scale=0.4,clip]{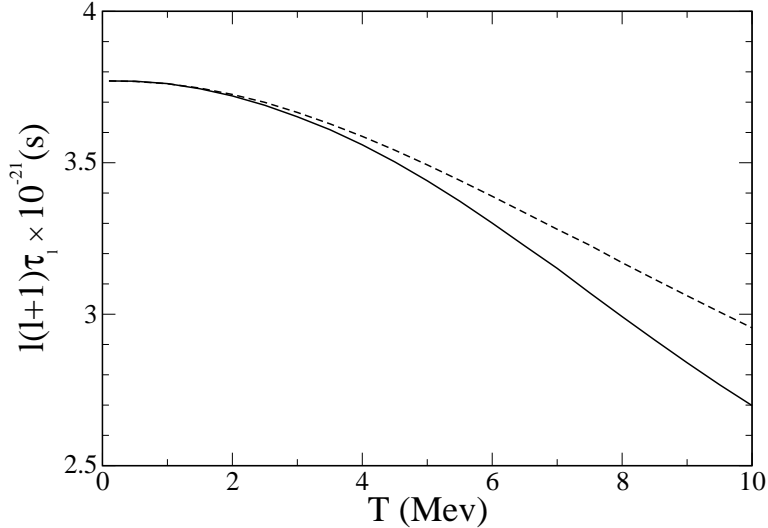}
\end{center}
\caption{Dependence of the renormalized relaxation time $l(l+1)\tau_l$ (\ref{taul}) 
on the temperature $T$. Solid curve is calculation by Eq. (\ref{taul1}), 
the dotted line is obtained in the low temperature approximation according to Eq. (\ref{taulf}).}
\label{fig1}
\end{figure}
The solid curve in \figurename~\ref{fig1} shows the calculation result of the renormalized relaxation 
time $l(l+1)\tau_l$ according to Eq. (\ref{taul1}) versus the temperature parameter $T$.
As you can see, with an increase in the temperature parameter, the renormalized relaxation time decreases in inverse 
proportion to the square of the temperature $\sim T^{-2}$. With a further increase in temperature 
the decrease in the renormalized relaxation time is close to linear.
This dependence is explained by the fact that when the Fermi system is heated the Fermi surface in momentum space
smears out and internucleon scattering becomes possible.
With increasing temperature the mean free path of nucleons decreases and the frequency of two-nucleon
collisions, which determines the intensity of dissipation of the collective excitation energy,
increases. 
The value of the energy dissipation intensity of the collective motion is inversely proportional 
to the relaxation time, therefore $\tau_l$ decreases with increasing $T$.
It should be noted that as a result of deformation of the Fermi surface even at zero temperature
the intensity of dissipation of the collective motion energy is nonzero.
Therefore, for $T=0$ the relaxation time $\tau_l$ has a finite value and in our case it is approximately equal 
to $3.77\cdot 10^{-21}$ s.

For comparison, we have calculated $l(l+1)\tau_l$ in the low-temperature approximation, 
in accordance with Eq. (\ref{taulf}). This type of calculation is indicated by a dashed curve.
The calculation of $l(l+1)\tau_l$ in the low temperature approximation (\ref{taulf}) practically coincides 
with the exact result (\ref{taul1}) up to $\sim 3$ MeV, where differences begin to appear.

Finally, it should be noted that the equation (\ref{dp8x}) was obtained without considering memory effects.
Therefore, the manifestation of these effects in the light of this approach requires further study and 
comparison with the previously obtained known results.

\section{Conclusions}

In this paper, we calculate the two-particle relaxation time of multipole collective excitations 
in the heated Fermi system within the framework of the diffusion approximation of the quantum kinetic theory.
The two-particle relaxation time is determined by the second moment in momentum space of the collision integral, 
which in the diffusion approximation is expressed in terms of the kinetic diffusion and drift coefficients.
In previous studies, these values have been approximated by constant coefficients and rough estimates have been made.

In our study, we use our earlier general expression for the diffusion coefficient in momentum space
and calculate the second moment of the collision integral in the general case of the momentum-dependent
kinetic diffusion and drift coefficients.
The obtained expression of the two-particle relaxation time depends on the temperature $T$.
In the case of low temperatures $T\ll \varepsilon_F$ the expression for the two-particle relaxation time 
transforms into the expression in the approximation of constant coefficients \cite{KoLu.UPJ.2014}.
Thus, the approximation of constant diffusion and drift coefficients does not exclude their temperature 
dependence.

To exclude from the analysis the Fermi surface distortion multipolarity $l$, 
the renormalized relaxation time $l(l+1)\tau_l$ is considered.
Numerical calculations have shown that in the low-temperature region, with an increase in $ T $, 
the value of $l(l+1)\tau_l$ decreases inversely quadratically $\sim T^{-2}$.
At high temperatures, the quadratic dependence becomes close to linear.
The calculation of $l(l+1)\tau_l$ in the low-temperature approximation (\ref{taulf}) is almost coincident 
with the exact result (\ref{taul1}) up to $\sim 3$ MeV, where differences start to appear.
Due to the multipole deformation of the Fermi surface, even at zero temperature, 
there is the relaxation of the collective motion of nucleons in the nucleus 
and the corresponding renormalized relaxation time $l(l+1)\tau_l$ has a finite 
value, which in our case is approximately equal to $3.77\cdot 10^{-21}$ s.

Since the expression for the transport diffusion coefficient (\ref{dp8x}) does not take into account 
the memory effects their manifestation requires further study.

It should also be noted that in contrast to the case of the collective excitations in the Fermi system
excitation of the particle-hole type is possible. 
The relaxation time of the particle-hole excitation $\tau_r$ depends on both 
transport coefficients $D_n$ and $K_n$ \cite{Wo.PRL.1982,KoLu.IJMP.2015}.
In the approximation of constant diffusion and drift coefficients, since $D_n$ and $K_n$ are related by Eq. (\ref{temp}), 
with an increase in the equilibrium temperature $T$ the value of $\tau_r$ increases \cite{Wo.PRL.1982} in contrast to 
the relaxation time of the collective multipole excitations $\tau_l$. 
In the general case of the momentum-dependent diffusion and drift coefficients, $\tau_r$ is determined by relaxation time
of the root-mean-square deviation of the distribution function from the initial to the equilibrium limit \cite{KoLu.IJMP.2015}.
The temperature dependence of $\tau_r$ in this case also requires further study.

\section{Acknowledgments}

This work was supported by the budget program "Support for the development 
of priority areas of scientific research” of the National Academy of Sciences of Ukraine
(Code 6541230, No. 0120U100434)".

\appendix\section{Calculation of integrals}

For $n=1$ we have
\begin{equation}
J_1=\int d\vec{p}\ Y_{lm}(\Omega_p) K_p \left(\frac{p}{m}\ \frac{\partial}{\partial p} 
[(1-2f_{eq})\delta f] + \frac{3}{m} (1-2f_{eq})\delta f \right).
\end{equation}
After substituting Eq. (\ref{dfsumynm}), we obtain
\begin{eqnarray}
&J_1&=-\sum_{l'm'}\nu_{l'm'}\int d\Omega_p\ Y_{lm}(\Omega_p) Y_{l'm'}(\Omega_p)
\nonumber \\
&\times& \int_0^\infty p^2dp\ K_p \left(\frac{p}{m}\ \frac{\partial}{\partial p} 
\left[(1-2f_{eq})\frac{m}{p}\frac{\partial f_{eq}}{\partial p}\right] 
+ \frac{3}{m} (1-2f_{eq})\frac{m}{p}\frac{\partial f_{eq}}{\partial p}\right),
\end{eqnarray}
where we took into account that
$$
\frac{\partial f_{eq}}{\partial\varepsilon}=\frac{m}{p}\frac{\partial f_{eq}}{\partial p}.
$$
Applying the relation
\begin{equation}
\int d\Omega_p \ Y_{lm}(\Omega_p) Y_{l'm'}(\Omega_p)=\delta_{ll'}\delta_{mm'},
\label{yynm}
\end{equation}
we obtain
\begin{equation}
J_1= - \nu_{lm} \int_0^\infty p^2dp\ K_p \left(p\ \frac{\partial}{\partial p} 
\left[(1-2f_{eq})\frac{1}{p}\frac{\partial f_{eq}}{\partial p}\right] 
+ \frac{3}{p} (1-2f_{eq})\frac{\partial f_{eq}}{\partial p}\right).
\end{equation}
We integrate the first term by parts, then we have
\begin{eqnarray}
J_1=&-&\nu_{lm}\left.\left\{p^2 K_p (1-2f_{eq})\frac{\partial f_{eq}}{\partial p} \right\vert_0^\infty 
-\int_0^\infty dp\ (1-2f_{eq})\left[3pK_p + p^2\frac{\partial K}{\partial p}\right] 
\frac{\partial f_{eq}}{\partial p} \right.
\nonumber \\
&+& \left. 3\int_0^\infty dp\ p K_p (1-2f_{eq})\frac{\partial f_{eq}}{\partial p} \right\}.
\end{eqnarray}
The first term in curly braces is zero, so after cancellation we get
\begin{equation}
J_1=\nu_{lm}\int_0^\infty dp\ p^2(1-2f_{eq})\frac{\partial K}{\partial p} 
\frac{\partial f_{eq}}{\partial p}.
\end{equation}

For $n=2$ we have
\begin{equation}
J_2 = \int d\vec{p}\ Y_{lm}(\Omega_p) (1-2f_{eq}) \delta f \frac{p}{m} \nabla_{p} K_p,
\end{equation}
or
\begin{eqnarray}
J_2 = & - &\sum_{l'm'}\nu_{l'm'} \int d\Omega_p \ Y_{lm}(\Omega_p) Y_{l'm'}(\Omega_p)
\nonumber \\
&\times & \int_0^\infty dp\ p^2 (1-2f_{eq}) \frac{m}{p} 
\frac{\partial f_{eq}}{\partial p} \frac{p}{m} \frac{\partial K_p}{\partial p}.
\end{eqnarray}
And after applying the relation (\ref{yynm}), we finally get
\begin{equation}
J_2 = - \nu_{lm}  \int_0^\infty dp\ p^2 (1-2f_{eq}) \frac{\partial f_{eq}}{\partial p} 
\frac{\partial K_p}{\partial p}.
\end{equation}

Similarly, for $n=3$, we obtain
\begin{eqnarray}
J_3&=&\int d\vec{p}\ Y_{lm}(\Omega_p) \delta f \nabla_{p} f_{eq} \cdot \nabla_{p} D_p
\nonumber \\
&=&-\sum_{l'm'}\nu_{l'm'} \int d\Omega_p \ Y_{lm}(\Omega_p) Y_{l'm'}(\Omega_p)
\int_0^\infty dp\ p^2 \frac{m}{p} \frac{\partial f_{eq}}{\partial p} \frac{\partial f_{eq}}{\partial p} 
\frac{\partial D_p}{\partial p}
\nonumber \\
&=&-m\nu_{lm}
\int_0^\infty dp\ p \left(\frac{\partial f_{eq}}{\partial p}\right)^2 \frac{\partial D_p}{\partial p}.
\end{eqnarray}

For $n=4$ we have
\begin{eqnarray}
J_4&=&\int d\vec{p}\ Y_{lm}(\Omega_p) (1-f_{eq}) \nabla_{p} \delta f \cdot \nabla_{p} D_p
\nonumber \\
&=& - \sum_{l'm'}\nu_{l'm'} 
\int d\Omega_p \ Y_{lm}(\Omega_p) Y_{l'm'}(\Omega_p)
\int_0^\infty dp\ p^2 (1-f_{eq}) \frac{\partial }{\partial p} \left(\frac{\partial f_{eq}}{\partial \varepsilon}\right) 
\frac{\partial D_p}{\partial p}
\nonumber \\
&=& -\nu_{lm} \int_0^\infty dp\ p^2 (1-f_{eq}) \frac{\partial }{\partial p} \left(\frac{\partial f_{eq}}{\partial \varepsilon}\right) 
\frac{\partial D_p}{\partial p}.
\end{eqnarray}
After integration by parts
\begin{eqnarray}
J_4=&-&\nu_{lm}\left\{ \left.p^2 (1-f_{eq}) \frac{\partial f_{eq}}{\partial \varepsilon} 
\frac{\partial D_p}{\partial p}\right\vert_0^\infty \right.
\nonumber \\
&-& \left.\int_0^\infty dp\ \frac{\partial f_{eq}}{\partial \varepsilon} \frac{\partial }{\partial p} 
\left(p^2 (1-f_{eq}) \frac{\partial D_p}{\partial p} \right) \right\}.
\end{eqnarray}
The first term is equal to zero, and in the second we write the derivatives
\begin{eqnarray}
J_4=&\nu_{lm}&\int_0^\infty dp\ \frac{m}{p} \frac{\partial f_{eq}}{\partial p} 
\left(2p (1-f_{eq}) \frac{\partial D_p}{\partial p} - p^2 \frac{\partial f_{eq}}{\partial p}
\frac{\partial D_p}{\partial p}\right.
\nonumber \\
&+&\left. p^2 (1-f_{eq}) \frac{\partial^2 D_p}{\partial p^2} \right).
\end{eqnarray}
So, we have
\begin{eqnarray}
J_4=&m&\nu_{lm}\left(2\int_0^\infty dp\ (1-f_{eq}) \frac{\partial f_{eq}}{\partial p} \frac{\partial D_p}{\partial p} 
- \int_0^\infty dp\ p \left(\frac{\partial f_{eq}}{\partial p}\right)^2 \frac{\partial D_p}{\partial p} \right.
\nonumber \\
&+&\left.\int_0^\infty dp\ p (1-f_{eq}) \frac{\partial f_{eq}}{\partial p} \frac{\partial^2 D_p}{\partial p^2} \right).
\end{eqnarray}

For $n=5$ we have
\begin{equation}
J_5=\int d\vec{p}\ Y_{lm}(\Omega_p) (1-2f_{eq}) \delta f \nabla^2_{p} D_p.
\end{equation}
We use the relation for the Laplace operator in a spherical coordinate system
$$
\nabla^2_{p}=\frac{1}{p^2} \frac{\partial}{\partial p} p^2\frac{\partial}{\partial p}
+\frac{1}{p^2\sin\theta} \frac{\partial}{\partial\theta} \sin\theta\frac{\partial}{\partial\theta}
+\frac{1}{p^2\sin^2\theta}\frac{\partial^2}{\partial\phi^2}.
$$
The expression for the diffusion coefficient $D_p$ (\ref{dp8x}) includes the distribution function, 
which we will consider as equilibrium and, therefore, spherically symmetric.
Thus, only the radial component will remain in the Laplacian
\begin{eqnarray}
J_5=&-&\sum_{l'm'}\nu_{l'm'} 
\int d\Omega_p \ Y_{lm}(\Omega_p) Y_{l'm'}(\Omega_p)
\nonumber \\
&\times &\int_0^\infty dp\ p^2 (1-2f_{eq}) \frac{m}{p} \frac{\partial f_{eq}}{\partial p} 
\frac{1}{p^2} \frac{\partial}{\partial p} \left(p^2\frac{\partial D_p}{\partial p}\right),
\end{eqnarray}
and after simplification we finally have
\begin{equation}
J_5= -m\nu_{lm} \int_0^\infty dp\ (1-2f_{eq}) \frac{\partial f_{eq}}{\partial p} 
\left(2\frac{\partial D_p}{\partial p}+p\frac{\partial^2 D_p}{\partial p^2}\right)
\end{equation}

For $n=6$ we have
\begin{equation}
J_6=\int d\vec{p}\ Y_{lm}(\Omega_p)  D_p \nabla^2_{p} \delta f,
\end{equation}
and
\begin{eqnarray}
J_6=&-&\sum_{l'm'}\nu_{l'm'} \left\{
\int d\Omega_p \ Y_{lm}(\Omega_p) Y_{l'm'}(\Omega_p)
\int_0^\infty dp\ p^2 D_p \frac{1}{p^2} \frac{\partial}{\partial p}\left( p^2\frac{\partial}{\partial p} 
\left(\frac{\partial f_{eq}}{\partial\varepsilon}\right)\right)\right. \nonumber \\
&+&\left. \int_0^\infty dp\ p^2 D_p \frac{\partial f_{eq}}{\partial\varepsilon} 
\int d\Omega_p \ Y_{lm}(\Omega_p) \frac{1}{p^2} \hat{\nabla}^2_p Y_{l'm'}(\Omega_p) 
\right\}.
\end{eqnarray}
Using the orthogonality relation for spherical functions
$$
\hat{\nabla}^2_p\ Y_{l'm'}(\Omega_p) = - l'(l'+1) Y_{l'm'}(\Omega_p),
$$
we write 
\begin{equation}
J_6=-\nu_{lm} \left\{
\int_0^\infty dp\ D_p \frac{\partial}{\partial p}\left( p^2\frac{\partial}{\partial p} \left(\frac{\partial f_{eq}}{\partial\varepsilon}\right)\right)
-l(l+1)\int_0^\infty dp\ D_p \frac{\partial f_{eq}}{\partial\varepsilon} \right\}.
\end{equation}
We integrate the first term by parts
\begin{eqnarray}
J_6=&-&\nu_{lm} \left\{ 
\left. D_p\ p^2 \frac{\partial}{\partial p} \left(\frac{\partial f_{eq}}{\partial\varepsilon}\right)\right\vert_0^\infty
-\int_0^\infty dp\ p^2 \frac{\partial}{\partial p} \left(\frac{\partial f_{eq}}{\partial\varepsilon}\right) 
\frac{\partial D_p}{\partial p} \right. \nonumber \\
&-&\left. l(l+1)m\int_0^\infty dp\ \frac{D_p}{p} \frac{\partial f_{eq}}{\partial p} \right\}.
\end{eqnarray}
The first term is equal to zero, and the second is integrated by parts
\begin{eqnarray}
J_6=&-&\nu_{lm} \left\{ 
-\left. p^2 \frac{\partial f_{eq}}{\partial\varepsilon} \frac{\partial D_p}{\partial p}\right\vert_0^\infty
+\int_0^\infty dp\ \frac{\partial f_{eq}}{\partial\varepsilon} \frac{\partial}{\partial p} 
\left( p^2 \frac{\partial D_p}{\partial p}\right) \right. \nonumber \\
&-&\left.l(l+1)m\int_0^\infty dp\ \frac{D_p}{p} \frac{\partial f_{eq}}{\partial p} \right\}.
\end{eqnarray}
The first term is again equal to zero, and in the second term we write out the derivative
\begin{eqnarray}
J_6=&-&\nu_{lm} \left\{ \int_0^\infty dp\ \frac{m}{p}\frac{\partial f_{eq}}{\partial p} 
\left( 2p \frac{\partial D_p}{\partial p}+p^2\frac{\partial^2 D_p}{\partial p^2}\right)\right. 
\nonumber \\
&-&\left.l(l+1)m\int_0^\infty dp\ \frac{D_p}{p} \frac{\partial f_{eq}}{\partial p} \right\}.
\end{eqnarray}
And finally we have
\begin{equation}
J_6= m\nu_{lm} \int_0^\infty dp\ \frac{\partial f_{eq}}{\partial p} \left( l(l+1) \frac{D_p}{p} 
- 2 \frac{\partial D_p}{\partial p} - p\frac{\partial^2 D_p}{\partial p^2}\right).
\end{equation}

\end{document}